\documentstyle[prl,aps]{revtex}
\begin{document}
\draft
\preprint{Dortmund, January 1994}

%  for the \twocolum option
%\advance\hsize0.5truecm\advance\hoffset-0.25truecm
%\advance\columnsep-0.5truecm\twocolumn[

\title{Spin-charge separation
       at small lengthscales
       in the 2D $t-J$ model}
\author{Claudius Gros and Roser Valent\'\i
       }
\address{Institut f\"ur Physik, Universit\"at Dortmund,
         44221 Dortmund, Germany
        }
\date{\today}
\maketitle
\begin{abstract}

%  for the \twocolum option
%\hbox to \hsize{\hfill\vbox{\hsize=14truecm

We consider projected wavefunctions for the 2D $t-J$ model.
For various wavefunctions, including correlated Fermi-liquid
and Luttinger-type wavefunctions we present the
static charge-charge and spin-spin structure factors.
Comparison with recent results from a high-temperature
expansion by Putikka {\it et al.} indicates spin-charge
separation at small lengthscales.

%  for the \twocolum option
%             }\hfill}
\end{abstract}
\pacs{71.45.Gm,77.70.Dm,74.70.Vy}
%%%%%%%%%%%%%%%%%%%%%%%%%%%%%%%%%%%%%%%%%%%
%%%% end bracket of \twocolumne option %%%%
%%%%%%%%%%%%%%%%%%%%%%%%%%%%%%%%%%%%%%%%%%%
%\maketitle
%]

%
%%%%%%%%%%%%%%%%%%%%%%%%%%%%%%%%%%%%%%%%%%%%%%%%%%%%%%%
%
%%%%%%%%%%%%%%%%%%%%%%%%%%%%%%%%%%%%%%%%%%%%%%%%%%%%%%%
%
%\narrowtext
\advance\baselineskip by 5pt
\renewcommand{\arraystretch}{2.0} \jot=8pt

%%%%% the * suppresses automatic numbering
\section*{Introduction}

The properties of correlated electrons in
two-dimensions (2D) are controversially discussed.
Anderson \cite{Anderson} has proposed that the well
established Luttinger-liquid properties
of one dimensional electron systems \cite{Haldane}
might occur, to
a certain extend, also in 2D. In 1D the spin and the
charge excitations move independently at small frequencies
\cite{Solyom},
a phenomena called {\it spin-charge separation}.
Separation of the spin and the charge degrees of freedom
has some interesting consequences on the
properties of the ground-state wavefunction.
In the framework of the $t-J$ model the charge
carriers are given by the holes which form in 1D,
since they move independently with respect
to the spin background,
a {\it hole Fermi-surface} distinct from the
Fermi-surface of the spins. This notion has been
shown to be rigorous in the $U/t\rightarrow\infty$ limit
of the 1D Hubbard model \cite{Ogata_Shiba}.

A huge amount of studies have been devoted, in the last
years, to the enlargement of our knowledge upon
2D correlated electron systems. It is probably fair to
say, that at the moment there is no method
in sight capable of establishing rigorously the properties of
2D correlated electron systems at large lengthscales,
or, what is equivalent, at small
differential wavevectors as it would be necessary
for the determination of true spin-charge separation in 2D.
It is, on the other hand,
an interesting question to ask, whether at small to
intermediate lengthscales spin and charge excitations can
move independently or not.

Considering a variational wavefunction originally introduced
by Hellberg and Mele in the context
of the 1D $t-J$ model \cite{Hellberg_Mele},
it has been shown \cite{VG}
that (i) it is possible to define Luttinger liquids in
two dimensions {\it variationally} and (ii) that the
projected kinetic energy favours Luttinger-liquid type
correlations in 2D, as it does in 1D \cite{Hellberg_Mele}.
Chen and Lee \cite{Chen_Lee} have shown that
the static spin-spin structure factor, $S_{\bf q}$, of this
Luttinger-liquid wavefunction compares favourably with
the $S_{\bf q}$ of the numerically obtained
exact ground state of a $10\times10$ cluster. Also, the recent
high-temperature expansion study by Putikka {\it et al.}
\cite{Putikka} has provided reliable data on $S_{\bf q}$ and
the static charge-charge structure factor, $N_{\bf q}$,
of the 2D $t-J$ model. An unexpected result has been the
observation of a prominent enhancement of $N_{\bf q}$ near
$(\pi,\pi)$ at quarter filling. Putikka {\it et al.}
found additionally that $N_{\bf q}$ could be fitted
very closely at all
densities with the curve appropiate for spinless Fermions,
i.e. assuming an independent hole Fermi-surface.
These results for $N_{\bf q}$ have been confirmed
at small densities by
a study using the power method, Quantum Monte Carlo
and a perturbation expansion \cite{Chen}.

The high-temperature expansion \cite{Putikka} and the
numerical studies \cite{Chen} can, of course,
not determine the properties of the 2D $t-J$ model
at long lengthscales and therefore address the
question of true spin-charge separation. They do,
on the other hand, provide reliable data on $S_{\bf q}$
and $N_{\bf q}$ at short to intermediate lengthscales.
Here we will interpret these data for the
correlation functions of the 2D $t-J$ model within
the context of projected wavefunctions. For a class of
variational wavefunctions with both short- and
long-ranged correlations we calculate
$S_{\bf q}$ and $N_{\bf q}$ and find that short-ranged
charge-charge repulsion can explain the data
by Putikka {\it et al.} and by Chen {\it et al.}
\cite{Putikka,Chen}.
Furthermore we show that it is the projected
kinetic energy which gives rise to the prominent
peak in $N_{\bf q}$.
Interpreting these results we come to the conclusion
that the strength of the short-ranged correlations
can only be explained when the the spin and the
charge degrees of freedom move independently at
small lengthscales. These results may partially
support Anderson's concept \cite{Anderson_confinement}
of incoherent one-particle tunneling in between the different
CuO-layers of the high-temperature superconductors.

%
%%%%%%%%%%%%%%%%%%%%%%%%%%%%%%%%%%%%%%%%%%%%%%%%%%%%%%%
%
%%%%%%%%%%%%%%%%%%%%%%%%%%%%%%%%%%%%%%%%%%%%%%%%%%%%%%%
%

\section*{Wavefunction}

We consider variational wavefunctions for the
2D $t-J$ model, which is given by
\begin{equation}
H_{t-J}\ =\ -t \sum_{\langle i,j\rangle,\sigma}
            ( \hat c_{j,\sigma}^{\dagger}
              \hat c_{i,\sigma}^{\phantom{\dagger}}
            + \hat c_{i,\sigma}^{\dagger}
              \hat c_{j,\sigma}^{\phantom{\dagger}} )
       +  J \sum_{\langle i,j\rangle}
            \hat {\bf S}_{i}\cdot \hat {\bf S}_{j},
\label{t_J}
\end{equation}
where the $\hat c_{i,\sigma}^{\dagger}$
($\hat c_{i,\sigma}^{\phantom{\dagger}}$)
are the creation (anihilation) operators on site $i$ of
electrons with spin $\sigma=\uparrow,\downarrow$, in the
subspace of no double occupancy. The
$\hat {\bf S}_{i}$ are the spin operators on site $i$ and
$\langle i,j\rangle$ denotes pairs of n.n. on the square
lattice.
The Jastrow-Luttinger wavefunction is given by \cite{VG}
\begin{equation}
|\Psi(T,S)\rangle\ =\
\exp\left[{-1\over 2T}\sum_{i<j} f({\bf r}_i-{\bf r}_j)
\,\hat n_i\,\hat n_j
     \right]
                  P_0\,|\psi_0\rangle,
\label{wavefunction}
\end{equation}
where $|\psi_0\rangle$ denotes the filled Fermi sea of electrons,
$P_0$ the projection operator on the subspace of no
double occupancy.
$
\hat n_i =
\hat c_{i,\uparrow}^{\dagger}
\hat c_{i,\uparrow}^{\phantom{\dagger}} +
\hat c_{i,\downarrow}^{\dagger}
\hat c_{i,\downarrow}^{\phantom{\dagger}}
$
is the density operator on site $i$ and
\begin{equation}
f({\bf r}_i-{\bf r}_j)\ = \
   -(1-S)
   (1-\delta_{\langle i,j\rangle})
\ln |{\bf r}_i-{\bf r}_j|
+S\,\delta_{\langle i,j\rangle}.
\label{Jastrow}
\end{equation}
In the limit $T\rightarrow\infty$ the wavefunction
defined by
Eq.\ (\ref{wavefunction})
reduces to the Gutzwiller wavefunction,
$P_0\,|\psi_0\rangle$. A finite $1/(2T)>0$ controls then
the strength of the repulsive \cite{note_1}
density-density Jastrow factor, given by
Eq.\ (\ref{Jastrow}).
The second variational parameter, $S$, controls the
details of the Jastrow factor. For $0\le S<1$ the
density-density correlation is long-ranged
\cite{note_2} and it
has been shown \cite{Hellberg_Mele,VG} that this logarithmic
correlator leads to an algebraic singularity in the
momentum distribution function at the Fermi-edge and
therefore to a Luttinger-liquid state. For $S=1$, on the
other side, the Jastrow correlator is short-ranged,
with only nearest neighbor repulsion (for $T>0$), and
the resulting state is a
correlated Fermi-liquid state \cite{VG}.

The Jastrow-Gutzwiller wavefunction, as defined by
Eq.\ (\ref{wavefunction}),
cannot be evaluated analytically. One then considers
large but finite clusters and calculates properties of
$|\Psi(T,S)\rangle$ numerically by the
variational Monte-Carlo method \cite{Annals}.
Here we are interested in the
static spin-spin structure factor,
\begin{equation}
S_{\bf q}\ =\
1/L \sum_{i,j}
e^{i\,{\bf q}({\bf r}_i-{\bf r}_j)}
<\hat S_i^z\,\hat S_j^z>,
\label{S_q}
\end{equation}
and the static charge-charge structure factor,
\begin{equation}
N_{\bf q}\ =\
1/L \sum_{i,j}
e^{i\,{\bf q}({\bf r}_i-{\bf r}_j)}
<\Delta\hat n_i\,\Delta\hat n_j>,
\label{N_q}
\end{equation}
where $L$ is the number of sites of the lattice
considered and $\Delta\hat n_i\equiv \hat n_i - n$, with
$n$ being the particle density.
In order to minimize finite-size effects
in cluster calculations it is advantageous
to consider closed shell configurations \cite{Annals}.
Here we consider clusters
with antiperiodic boundary conditions
which tile the square lattice in
${\bf L}_1 =(2m,2m-2)$  and
${\bf L}_2 =(-2m+2,2m)$ directions (with $m$ an integer).
In Fig.\ (\ref{lattice})
we illustrate the first
Brillouin zone for a lattice with $m=6$, which has
$L=12^2+10^2=244$ sites.  Also shown in
Fig.\ (\ref{lattice}) are the electron
Fermi-sea (filled circles) and the hole
Fermi-sea (filled squares) for $N_h=124$ holes.
This type of lattices has the property that,
whenever the number of holes is a multiple of
four, but not of eight, both the electron
Fermi-sea {\it and} the hole Fermi-sea
are closed-shell configurations, as illustrated in
Fig.\ (\ref{lattice}). The dashed lines in
Fig.\ (\ref{lattice}) denote the locations of the
Fermi surfaces for the same nominal densities
$n=(L-N_h)/L$ of electrons and of
$N_h/L$ holes respectively,
in the thermodynamic limit. Here the
Fermi surfaces in the thermodynamic limit are
consistent with the respective Fermi surfaces of the
finite cluster
(this is not necessarily always the case
\cite{VG}).

%
%%%%%%%%%%%%%%%%%%%%%%%%%%%%%%%%%%%%%%%%%%%%%%%%%%%%%%%
%
%%%%%%%%%%%%%%%%%%%%%%%%%%%%%%%%%%%%%%%%%%%%%%%%%%%%%%%
%

\section*{Results}

In
Fig.\ (\ref{bill_1})
we present for
various {\bf q} in the first Brillouin zone
the static charge-charge structure factor,
$N_{\bf q}$, as defined by
Eq.\ (\ref{N_q}), for correlated Fermi-liquid
wavefunctions,
$|\Psi(T,S=1)\rangle$ as defined by
Eq.\ (\ref{wavefunction}), for the lattice
illustrated in
Fig.\ (\ref{lattice})
with $L=244$ sites and
$N_h=124$ corresponding to a particle
density $n=(L-N_h)/L\approx0.49$.
In
Fig.\ (\ref{bill_1})
we have included for comparison the curve for
$N_{\bf q}$ for spinless Fermions, i.e. assuming
a well defined hole Fermi-sea as illustrated in
Fig.\ (\ref{lattice}).
We observe that data for the Gutzwiller state, realized
for $1/(2T)=0$, is qualitatively different from the
curve for spinless Fermions. The data for
$|\Psi(1/(2T)=0.2,S=1)\rangle$ is, on the other
hand very close to the curve for spinless Fermions and
agree remarkable well with the results of
Putikka {\it et al.} and Chen {\it et al.} \cite{Putikka,Chen}.
Noting that the correlated Fermi-liquid wavefunction
$|\Psi(1/(2T)=0.2,S=1)\rangle$ contains explicitly
only nearest neighbour correlations we conclude that
the static density-density correlation factor of the
2D $t-J$ model is dominated by short-distance effects. This
does, of course, not exclude the possibility of long-range
correlations to be present in the paramagnetic state of the
2D $t-J$ model, as these correlations cannot be calculated
rigorously with present-day methods.

The energetics behind the results for
$N_{\bf q}$ is dominated by the projected kinetic
energy for small $J/t$. The exchange hole between
parallel spins in the Gutzwiller wavefunction is consequence
of the Fermi statistics. In the $t-J$ model the motion of
electrons with spin $\sigma$ is blocked also by the
electrons with spin $-\sigma$. A favourable kinetic energy
is then obtained in a state in which an additional exchange
hole in between electrons with opposite spins is simulated. This is
exactly what the Jastrow factor defined by
Eq.\ (\ref{Jastrow}) does, as it does contain short-ranged
repulsion between all electrons, regardless of their relative
spin orientation. Consequently the probability of finding
two n.n. sites both occupied decreases with increasing $1/(2T)$
(for $S=1$) and so does $N_{(\pi,0)}$ as evidenced in
Fig.\ (\ref{bill_1}). At quarter filling there is one
particle for every second site, which leads to a large
increase of the probability of finding two n.n.n. sites
both occupied and consequently $N_{(\pi,\pi)}$ increases
with increasing $1/(2T)$.
The extreme limit $1/(2T)\rightarrow\infty$
is energetically not advantageous, as it corresponds
to a charge-density wave with a delta function
at $(\pi,\pi)$ and zero kinetic energy. Variationally
we find the optimal value for $1/(2T)\approx0.15$ (within the
constraint $S\equiv1$) for values of $J\le0.25 t$. Within the
accuracy of the variational calculations this agrees fairly
well with the $1/(2T)\approx0.2$ which would give the best
result for $N_{\bf q}$.

The fact that the system wants to simulate an
exchange hole between all electrons, regardless
of the relative spin orientation has a natural
explanation within the spinless Fermion model. In
the case the holes would form a distinct Fermi-sea,
as illustrated in
Fig.\ (\ref{lattice}), a exchange hole in between
electrons of both spin orientations would
be one of the consequences. Even a partial formation
of a hole Fermi-sea, i.e. with a washed-out Fermi-surface,
would also result in an effective, short-ranged
charge-charge repulsion, as observed in the data for
$N_{\bf q}$. Note that at quarter filling $\pi/k_F\approx 1$.
We therefore interpret the data presented
in Fig.\ (\ref{bill_1}) as evidence for
spin-charge separation at small lengthscales.

In Fig.\ (\ref{bill_2}) we present the results
at quarter filling for
$N_{\bf q}$ for the state with the best kinetic energy,
allowing for any value of $S$. The optimal state
is given by $1/(2T)\approx0.15,\ S\approx0.6$, which
is a Luttinger-liquid state \cite{VG}. It has
qualitatively the same features as the states
with n.n. correlations only, shown in
Fig.\ (\ref{bill_1}), but more washed out.
In Fig.\ (\ref{bill_3}) we present the results for
$N_{\bf q}$ for the Gutzwiller state for the cluster
with $L=244$ sites and $N_h=124$ and $N_h=52$ holes,
corresponding to electron densities
$n\approx0.49$ and $\approx0.79$ respectively.
The solid lines are the corresponding curves for
spinless Fermions. We see that near half-filling the
magnitude of $N_{\bf q}$, which again is very
close to that of the spinless fermions \cite{Putikka},
is reproduced by
the Gutzwiller wavefunction. The data points
for states $1/(2T)>0$ do not differ much from the
data of the Gutzwiller state at these fillings, as
density fluctuations are reduced by the projection
operator.

It is of interest to compare the variational Monte
Carlo data for the Gutzwiller wavefunction with
the predictions of the
Gutzwiller approximation formula (GAF),
which relates, by simple counting arguments
\cite{renormalized}, the $N_{\bf q}$ in
the projected state, as defined by
Eq.\  (\ref{N_q}), with the
$N_{\bf q}^{(0)}$ in the unprojeced state, defined by
\begin{equation}
N_{\bf q}^{(0)}\ =\
1/L \sum_{i,j}
e^{i\,{\bf q}({\bf r}_i-{\bf r}_j)}
\langle\psi_0|\Delta\hat n_i\,\Delta\hat n_j|\psi_0\rangle/
\langle\psi_0|\psi_0\rangle
\label{N_q_0}
\end{equation}
via the approximative formula
\begin{equation}
N_{\bf q}\ \approx\ {1-n\over1-n/2}
N_{\bf q}^{(0)}.
\label{GAF_N_q}
\end{equation}
We have included in Fig.\ (\ref{bill_3})
the predictions of the GAF, as given by
Eq.\ (\ref{GAF_N_q}) (dashed lines). The
overall magnitude of $N_{\bf q}$ is well
captured by the GAF, thought the variational
Monte Carlo data has the tendency to be
smoother at small dopings.
The deviation of the
data for the Gutzwiller state at $n\approx0.79$
from the corresponding curve for
spinless Fermions for small {\bf q}-vectors is
open to interpretation. Note, that no reliable results
do exist for $N_{\bf q}$ in this region
\cite{Putikka,Chen}. For the
static spin-spin structure factor the
Gutzwiller approximative fromula reads as
\begin{equation}
S_{\bf q}\ \approx\ {1\over1-n/2}
S_{\bf q}^{(0)}.
\label{GAF_S_q}
\end{equation}
In Fig.\ (\ref{bill_4}) we present
Eq.\ (\ref{GAF_S_q}) at quarter filling together
with the variational Monte Carlo data for
$S_{\bf q}$ for the Gutzwiller wavefunction
(filled triangles), the best Luttinger liquid
wavefunction (filled circles) and the
Fermi-liquid wavefunction with n.n. correlations
only (filled squares). The overall magnitude
is well reproduced by the GAF. Note that the
Luttinger liquid state would have in the thermodynamic
limit a cusp at $2{\bf k}_F$ with an algebraic
singularity (at quarter filling $2{\bf k}_F=(\pi,\pi)$
along the (1,1) direction).

In conclusion we have shown that the known features
of the static charge-charge structure factor of the
2D $t-J$ model are dominated by short-ranged correlations.
We interpretat the results as evidence for
spin-charge separation at small lengthscales.

We would like to thank Y.C. Chen, E. Dagotto,
W.O. Putikka and T.M. Rice for discussions.
This work was supported by the Deutsche Forschungsgemeinschaft,
by the Minister f\"ur Wissenschaft
und Forschung des Landes Nordrhein-Westfalen and by the
European Community European Strategic Program for
Research in Information Technology program, Project
No. 3041-MESH.
%
%%%%%%%%%%%%%%%%%%%%%%%%%%%%%%%%%%%%%%%%%%%%%%%%%%%%%%%
%
%%%%%%%%%%%%%%%%%%%%%%%%%%%%%%%%%%%%%%%%%%%%%%%%%%%%%%%
%

%
%%%%%%%%%%%%%%%%%%%%%%%%%%%%%%%%%%%%%%%%%%%%%%%%%%%%%%%
%
%%%%%%%%%%%%%%%%%%%%%%%%%%%%%%%%%%%%%%%%%%%%%%%%%%%%%%%
%
\begin{figure}
\caption{ An illustration of the Brillouin zone of the
          $12^2+10^2=244$ lattice with antiperiodic
          boundary conditions. The dashed lines indicate
          the location of the particle and the hole
          fermi surface at a filling
          $ n=(244-124)/244\approx0.49$. The full circles/squares
          denote the {\bf k}-states occupied by particles/holes
          on the finite lattice. Note that both the particle
          {\it and} the hole fermi-surfaces are closed-shell
          configurations.
          }
\label{lattice}
\end{figure}
\begin{figure}
\caption{ The charge-charge structure factor, $N_q$
          for various $q$-directions in the 2D Brillouin
          zone at quarter filling ($n=120/244$).
          The filled circles are the results
          for the Fermi-liquid wavefunction with
          n.n. density-density correlation in the Jastrow
          prefactor of strenght $1/(2T) = 0.0,\ 0.15\
          0.2$ and $0.25$ respectively. The full line is
          the $N_q$ of spinless fermions, the dotted
          lines are guides to the eye.
          }
\label{bill_1}
\end{figure}
\begin{figure}
\caption{ The charge-charge structure factor, $N_q$
          for various $q$-directions in the 2D Brillouin
          zone at quarter filling.
          The filled circles are the results of
          the optimal Luttinger-liquid wavefunction and the
          full line is the $N_q$ of spinless fermions,
          the dotted lines are guides to the eye.
          }
\label{bill_2}
\end{figure}
\begin{figure}
\caption{ The charge-charge structure factor, $N_q$
          for various $q$-directions in the 2D Brillouin
          zone for fillings
          $ n=(244-124)/244\approx0.49$ and
          $ n=(244-52)/244\approx0.79$.
          The filled circles are the results
          for the Gutzwiller wavefunction, the
          dashed line the predictions of the
          Gutzwiller approximation formula (GAF)
          and the full lines the respective
          $N_q$ of spinless fermions.
          The dotted lines are guides to the eye.
          }
\label{bill_3}
\end{figure}
\begin{figure}
\caption{ The spin-spin structure factor, $S_q$,
          for various $q$-directions in the 2D Brillouin
          zone at quarter filling.
          The results for the Gutzwiller wavefunction
          (triangles), for the optimal Luttinger-liquid
          wavefunction (circles) and for the correlated
          Fermi-liquid wavefunctions (squares) are compared
          with the Gutzwiller approximative formula
          (GAF), given by the dashed line.
          The dotted lines are guides to the eye.
          }
\label{bill_4}
\end{figure}

\begin{references}
\bibitem{note_2} A slight difference in the definition of the
                 Jastrow correlator, Eq.\ (\ref{Jastrow}), with
                 respect to Ref. \cite{VG} is the factor
                 $ (1-\delta_{\langle i,j\rangle})
                 $ in front of the $log$ term.
\end{references}
\end{document}